\documentclass[a4paper,12pt]{article}
\usepackage{graphicx}

\newcommand{\alphat}{\tilde\alpha}
\newcommand{\betat}{\tilde\beta}
\newcommand{\kappat}{\tilde\kappa}
\newcommand{\deltat}{\tilde\delta}
\newcommand{\epsilont}{\tilde\epsilon}

\begin{document}

\newcommand{\be}{\begin{equation}}
\newcommand{\beq}{\begin{equation}}
\newcommand{\eeq}{\end{equation}}
\newcommand{\ee}{\end{equation}}

\newcommand{\refeq}[1]{Eq.\ref{eq:#1}}
\newcommand{\refig}[1]{Fig.\ref{fig:#1}}
\newcommand{\refsec}[1]{Sec.\ref{sec:#1}}

\newcommand{\beqn}{\begin{eqnarray}}
\newcommand{\eeqn}{\end{eqnarray}}
\newcommand{\bea}{\begin{eqnarray}}
\newcommand{\ena}{\end{eqnarray}}
\newcommand{\ra}{\rightarrow}
\newcommand{\susy}{{{\cal SUSY}$\;$}}
\newcommand{\su}{$ SU(2) \times U(1)\,$}

\newcommand{\gag}{$\gamma \gamma$ }
\newcommand{\gagt}{\gamma \gamma }
\newcommand{\gam}{\gamma \gamma }
\def\W{{\mbox{\boldmath $W$}}}
\def\B{{\mbox{\boldmath $B$}}}
\def\V{{\mbox{\boldmath $V$}}}
\newcommand{\np}{Nucl.\,Phys.\,}
\newcommand{\pl}{Phys.\,Lett.\,}
\newcommand{\pr}{Phys.\,Rev.\,}
\newcommand{\prl}{Phys.\,Rev.\,Lett.\,}
\newcommand{\prep}{Phys.\,Rep.\,}
\newcommand{\zp}{Z.\,Phys.\,}
\newcommand{\sovjnp}{{\em Sov.\ J.\ Nucl.\ Phys.\ }}
\newcommand{\nuclinst}{{\em Nucl.\ Instrum.\ Meth.\ }}
\newcommand{\annp}{{\em Ann.\ Phys.\ }}
\newcommand{\intjmp}{{\em Int.\ J.\ of Mod.\  Phys.\ }}

\newcommand{\eps}{\epsilon}
\newcommand{\mw}{M_{W}}
\newcommand{\mww}{M_{W}^{2}}
\newcommand{\mwmw}{M_{W}^{2}}
\newcommand{\mhmh}{M_{H}^2}
\newcommand{\mz}{M_{Z}}
\newcommand{\mzz}{M_{Z}^{2}}

\newcommand{\cw}{\cos\theta_W}
\newcommand{\sw}{\sin\theta_W}
\newcommand{\tw}{\tan\theta_W}
\def\tww{\tan^2\theta_W}
\def\stw{s_{2w}}

\newcommand{\smw}{s_M^2}
\newcommand{\cmw}{c_M^2}
\newcommand{\seff}{s_{{\rm eff}}^2}
\newcommand{\ceff}{c_{{\rm eff}}^2}
\newcommand{\seffl}{s_{{\rm eff\;,l}}^{2}}
\newcommand{\sww}{s_W^2}
\newcommand{\cww}{c_W^2}
\newcommand{\swo}{s_W}
\newcommand{\cwo}{c_W}

\newcommand{\epm}{$e^{+} e^{-}\;$}
\newcommand{\epemt}{$e^{+} e^{-}\;$}
\newcommand{\epem}{e^{+} e^{-}\;}
\newcommand{\ememt}{$e^{-} e^{-}\;$}
\newcommand{\emem}{e^{-} e^{-}\;}

\newcommand{\lra}{\leftrightarrow}
\newcommand{\tr}{{\rm Tr}}
\def\ls1{{\not l}_1}
\newcommand{\cms}{centre-of-mass\hspace*{.1cm}}


\newcommand{\dkg}{\Delta \kappa_{\gamma}}
\newcommand{\dkz}{\Delta \kappa_{Z}}
\newcommand{\dz}{\delta_{Z}}
\newcommand{\dgz}{\Delta g^{1}_{Z}}
\newcommand{\dgzt}{$\Delta g^{1}_{Z}\;$}
\newcommand{\la}{\lambda}
\newcommand{\lag}{\lambda_{\gamma}}
\newcommand{\lambdae}{\lambda_{e}}
\newcommand{\laz}{\lambda_{Z}}
\newcommand{\lnl}{L_{9L}}
\newcommand{\lnr}{L_{9R}}
\newcommand{\lt}{L_{10}}
\newcommand{\lu}{L_{1}}
\newcommand{\ld}{L_{2}}
\newcommand{\eeww}{e^{+} e^{-} \ra W^+ W^- \;}
\newcommand{\eewwt}{$e^{+} e^{-} \ra W^+ W^- \;$}
\newcommand{\epemww}{e^{+} e^{-} \ra W^+ W^- }
\newcommand{\epemwwt}{$e^{+} e^{-} \ra W^+ W^- \;$}
\newcommand{\eennhht}{$e^{+} e^{-} \ra \nu_e \bar \nu_e HH\;$}
\newcommand{\eennhh}{e^{+} e^{-} \ra \nu_e \bar \nu_e HH\;}
\newcommand{\ppwg}{p p \ra W \gamma}
\newcommand{\wwhh}{W^+ W^- \ra HH\;}
\newcommand{\wwhht}{$W^+ W^- \ra HH\;$}
\newcommand{\ppwz}{pp \ra W Z}
\newcommand{\ppwgt}{$p p \ra W \gamma \;$}
\newcommand{\ppwzt}{$pp \ra W Z \;$}
\newcommand{\gamgamt}{$\gamma \gamma \;$}
\newcommand{\gamgam}{\gamma \gamma \;}
\newcommand{\egamt}{$e \gamma \;$}
\newcommand{\egam}{e \gamma \;}
\newcommand{\gamgamwwt}{$\gamma \gamma \ra W^+ W^- \;$}
\newcommand{\gamgamwwht}{$\gamma \gamma \ra W^+ W^- H \;$}
\newcommand{\gamgamwwh}{\gamma \gamma \ra W^+ W^- H \;}
\newcommand{\gamgamwwhht}{$\gamma \gamma \ra W^+ W^- H H\;$}
\newcommand{\gamgamwwhh}{\gamma \gamma \ra W^+ W^- H H\;}
\newcommand{\ggww}{\gamma \gamma \ra W^+ W^-}
\newcommand{\ggwwt}{$\gamma \gamma \ra W^+ W^- \;$}
\newcommand{\ggwwht}{$\gamma \gamma \ra W^+ W^- H \;$}
\newcommand{\ggwwh}{\gamma \gamma \ra W^+ W^- H \;}
\newcommand{\ggwwhht}{$\gamma \gamma \ra W^+ W^- H H\;$}
\newcommand{\ggwwhh}{\gamma \gamma \ra W^+ W^- H H\;}
\newcommand{\ggwwz}{\gamma \gamma \ra W^+ W^- Z\;}
\newcommand{\ggwwzt}{$\gamma \gamma \ra W^+ W^- Z\;$}

\newcommand{\veps}{\varepsilon}

\newcommand{\ptu}{p_{1\bot}}
\newcommand{\vecptu}{\vec{p}_{1\bot}}
\newcommand{\ptd}{p_{2\bot}}
\newcommand{\vecptd}{\vec{p}_{2\bot}}
\newcommand{\ie}{{\em i.e.}}
\newcommand{\cm}{{{\cal M}}}
\newcommand{\cl}{{{\cal L}}}
\newcommand{\cd}{{{\cal D}}}
\newcommand{\cv}{{{\cal V}}}
\def\slashc{c\kern -.400em {/}}
\def\slashp{p\kern -.400em {/}}
\def\slashL{L\kern -.450em {/}}
\def\slashcl{\cl\kern -.600em {/}}
\def\slashr{r\kern -.450em {/}}
\def\slashk{k\kern -.500em {/}}
\def\Ww{{\mbox{\boldmath $W$}}}
\def\B{{\mbox{\boldmath $B$}}}
\def\noi{\noindent}
\def\nn{\noindent}
\def\sm{${\cal{S}} {\cal{M}}\;$}
\def\smn{${\cal{S}} {\cal{M}}$}
\def\nph{${\cal{N}} {\cal{P}}\;$}
\def\sb{$ {\cal{S}}  {\cal{B}}\;$}
\def\ssb{${\cal{S}} {\cal{S}}  {\cal{B}}\;$}
\def\ssbe{{\cal{S}} {\cal{S}}  {\cal{B}}}
\def\cviol{${\cal{C}}\;$}
\def\pviol{${\cal{P}}\;$}
\def\cpviol{${\cal{C}} {\cal{P}}\;$}

\newcommand{\lgg}{\lambda_1\lambda_2}
\newcommand{\lww}{\lambda_3\lambda_4}
\newcommand{\ppin}{ P^+_{12}}
\newcommand{\pmin}{ P^-_{12}}
\newcommand{\ppout}{ P^+_{34}}
\newcommand{\pmout}{ P^-_{34}}
\newcommand{\sinsq}{\sin^2\theta}
\newcommand{\cossq}{\cos^2\theta}
\newcommand{\yt}{y_\theta}
\newcommand{\hppll}{++;00}
\newcommand{\hpmll}{+-;00}
\newcommand{\hpplt}{++;\lambda_30}
\newcommand{\hpmlt}{+-;\lambda_30}
\newcommand{\hpptt}{++;\lambda_3\lambda_4}
\newcommand{\hpmtt}{+-;\lambda_3\lambda_4}
\newcommand{\dk}{\Delta\kappa}
\newcommand{\klam}{\Delta\kappa \lambda_\gamma }
\newcommand{\kac}{\Delta\kappa^2 }
\newcommand{\lac}{\lambda_\gamma^2 }
\def\gamgamtzz{$\gamma \gamma \ra ZZ \;$}
\def\gamgamtww{$\gamma \gamma \ra W^+ W^-\;$}
\def\gamgamtwwe{\gamma \gamma \ra W^+ W^-}

\def\intfd{ \int \frac{d^4 r}{(2\pi)^4} }
\def\intnd{ \int \frac{d^n r}{(2\pi)^n} }
\def\intnmu{ \mu^{4-n} \int \frac{d^n r}{(2\pi)^n} }
\newcommand{\Dkm}{[(r+k)^2-m_2^2]}
\newcommand{\Dkom}{[(r+k_1)^2-m_2^2]}
\newcommand{\Dkotm}{[(r+k_1+k_2)^2-m_3^2]}
\def\piggt{$\Pi_{\gamma \gamma}\;$}
\def\pigg{\Pi_{\gamma \gamma}}
\newcommand{\mn}{{\mu \nu}}
\newcommand{\mzb}{M_{Z,0}}
\newcommand{\mzbs}{M_{Z,0}^2}
\newcommand{\mwb}{M_{W,0}}
\newcommand{\mwbs}{M_{W,0}^2}
\newcommand{\dgg}{\frac{\delta g^2}{g^2}}
\newcommand{\dee}{\frac{\delta e^2}{e^2}}
\newcommand{\dss}{\frac{\delta s^2}{s^2}}
\newcommand{\dmw}{\frac{\delta \mww}{\mww}}
\newcommand{\dmz}{\frac{\delta \mzz}{\mzz}}
\def\pigz{\Pi_{\gamma Z}}
\def\pizz{\Pi_{Z Z}}
\def\piww{\Pi_{WW}}
\def\pioo{\Pi_{11}}
\def\pitt{\Pi_{33}}
\def\pitq{\Pi_{3Q}}
\def\piqq{\Pi_{QQ}}
\def\delr{\Delta r}
\def\calm{{\cal {M}}}
\def\gww{G_{WW}}
\def\gzz{G_{ZZ}}
\def\goo{G_{11}}
\def\gtt{G_{33}}
\def\szz{s_Z^2}
\def\estk{e_\star^2(k^2)}
\def\sstk{s_\star^2(k^2)}
\def\cstk{c_\star^2(k^2)}
\def\sstz{s_\star^2(\mzz)}
\def\mzst{{M_Z^{\star}}(k^2)^2}
\def\mwst{{M_W^{\star}}(k^2)^2}
\def\epo{\varepsilon_1}
\def\epd{\varepsilon_2}
\def\ept{\varepsilon_3}
\def\dro{\Delta \rho}
\def\gmu{G_\mu}
\def\alpz{\alpha_Z}
\def\danpmz{\Delta\alpha_{{\rm NP}}(\mzz)}
\def\danpk{\Delta\alpha_{{\rm NP}}(k^2)}
\def\calt{{\cal {T}}}
\def\piggh{\pigg^h(s)}
\def\cuv{C_{UV}}
\def\pilr{G_{LR}}
\def\pill{G_{LL}}
\def\dak{\Delta \alpha(k^2)}
\def\damz{\Delta \alpha(\mzz)}
\def\dahmz{\Delta \alpha^{(5)}_{{\rm had}}(\mzz)}
\def\sth{s_{\theta}^2}
\def\cth{c_{\theta}^2}
\newcommand{\siki}[1]{Eq.\ref{eq:#1}}
\newcommand{\zu}[1]{Fig.\ref{fig:#1}}
\newcommand{\setu}[1]{Sec.\ref{sec:#1}}
\newcommand{\anlg}{\tilde\alpha}
\newcommand{\bnlg}{\tilde\beta}
\newcommand{\dnlg}{\tilde\delta}
\newcommand{\enlg}{\tilde\varepsilon}
\newcommand{\knlg}{\tilde\kappa}
\newcommand{\xiw}{\xi_W}
\newcommand{\xiz}{\xi_Z}
\newcommand{\dbr}{\delta_B}
\newcommand{\bothd}{{ \leftrightarrow \atop{\partial^{\mu}} } }

\newcommand{\BARE}[1]{\underline{#1}}
\newcommand{\ZF}[1]{\sqrt{Z}_{#1}}
\newcommand{\ZFT}[1]{\tilde{Z}_{#1}}
\newcommand{\ZH}[1]{\delta Z_{#1}^{1/2}}
\newcommand{\ZHb}[1]{\delta Z_{#1}^{1/2\,*}}
\newcommand{\DM}[1]{\delta M^2_{#1}}
\newcommand{\DMS}[1]{\delta M_{#1}}
\newcommand{\Dm}[1]{\delta m_{#1}}
\newcommand{\tree}[1]{\langle{#1}\rangle}

\newcommand{\Cuv}{C_{{\rm UV}}}
\newcommand{\logw}{\log M_W^2}
\newcommand{\logz}{\log M_Z^2}
\newcommand{\logh}{\log M_H^2}
\newcommand{\swt}{s_W^2}
\newcommand{\cwt}{c_W^2}
\newcommand{\swf}{s_W^4}
\newcommand{\cwf}{c_W^4}
\newcommand{\MWt}{M_W^2}
\newcommand{\MZt}{M_Z^2}
\newcommand{\MHt}{M_H^2}

\newcommand{\VECsl}[1]{\not{#1}}

\newcommand{\Bphi}{\mbox{\boldmath$\phi$}}

\begin{titlepage}
\def\baselinestretch{1.2}
\vspace*{\fill}
\begin{center}
{\large {\bf {\em Full one-loop electroweak radiative corrections
to single Higgs production in \epemt .}}}

\vspace*{.5cm}


G. B\'elanger${}^{1)}$, F. Boudjema${}^{1)}$, J.
Fujimoto${}^{2)}$, T. Ishikawa${}^{2)}$, \\ T. Kaneko${}^{2)}$, K.
Kato${}^{3)}$,  Y.
Shimizu${}^{2)}$ \\

\vspace{4mm}

{\it 1) LAPTH${\;}^\dagger$, B.P.110, Annecy-le-Vieux F-74941,
France.}
\\ {\it
2) KEK, Oho 1-1, Tsukuba, Ibaraki 305--0801, Japan.} \\
{\it 3) Kogakuin University, Nishi-Shinjuku 1-24, Shinjuku, Tokyo
163--8677, Japan.} \\

\vspace{10mm}

\end{center}

\centerline{ {\bf Abstract} } \baselineskip=14pt \noindent
{\small We present the full ${{\cal O}}(\alpha)$ electroweak
radiative corrections to single Higgs production in \epemt. This
takes into account the full one-loop corrections as well as the
effects of hard photon radiation. We include  both the fusion and
Higgs-strahlung processes. The computation is performed with the
help of {\tt GRACE-loop} where we have implemented a generalised
non-linear gauge fixing condition. The latter includes $5$ gauge
parameters that can be used for checks on our results. Besides the
UV, IR finiteness and gauge parameter independence checks it
proves also powerful to test our implementation of the 5-point
function. We find that for a 500GeV machine and a light Higgs of
mass $150$GeV, the total ${{\cal O}}(\alpha)$ correction is small
when the results are expressed in terms of $\alpha_{{\rm QED}}$.
The total correction decreases slightly for higher energies. For
moderate centre of mass energies the total ${{\cal O}}(\alpha)$
decreases as the Higgs mass increases, reaching $-10\%$ for
$M_H=350$GeV and $\sqrt{s}=500$GeV. In order to quantify the
genuine weak corrections we have  subtracted the universal virtual
and bremsstrahlung correction from the full ${{\cal O}}(\alpha)$.
We find, for $M_H=150$GeV, a weak correction slowly decreasing
from $-2\%$ to $-4\%$ as the energy increases from
$\sqrt{s}=300$GeV to $\sqrt{s}=1$TeV after expressing the
tree-level results in terms of $G_\mu$.}
\vspace*{\fill}

\vspace*{0.1cm} \rightline{LAPTH-959/2002}
\vspace*{0.1cm}\rightline{KEK-CP-135}

$^\dagger${\small UMR 5108 du CNRS, associ\'ee  \`a l'Universit\'e
de Savoie.} \normalsize
\end{titlepage}


\baselineskip=18pt

\setcounter{section}{0}
\setcounter{subsection}{0}
\setcounter{equation}{0} \setcounter{table}{0}
\def\thesubsection {\thesection.\arabic{subsection}}
\def\theequation{\thesection.\arabic{equation}}

\newcommand{\grc}{{\tt GRACE}$\;$}
\newcommand{\grcp}{{\tt GRACE}}
\newcommand{\grcl}{{\tt GRACE-loop$\;$}}
\newcommand{\nnhet}{$\epem \ra \nu_e \bar{\nu}_e H \;$}
\newcommand{\nnhe}{$\epem \ra \nu_e \bar{\nu}_e H$}
\newcommand{\nnht}{$\epem \ra \nu \bar{\nu} H \;$}
\newcommand{\nnh}{$\epem \ra \nu \bar{\nu} H$}
\newcommand{\eezh}{$\epem \ra Z H$}
\newcommand{\eezht}{$\epem \ra Z H \;$}
\section{Introduction}
Uncovering the mechanism of symmetry breaking is one of the major
tasks of the high energy colliders. Most prominent is the search
for the Higgs particle. Although the LHC should not miss this
particle even if it weighed up to 1TeV, precision measurements on
the Higgs properties will only be conducted in an \epemt collider.
There are two important mechanisms for Higgs production in \epemt.
The Higgs-strahlung process, \eezht and the $W$-fusion process,
\nnhet. The former is the dominant one at small (in the LEP2 range
say)  to moderate energies but decreases rather fast with energy.
At TeV energies the $W$-fusion process dominates by far for Higgs
masses up to $1$TeV. Even at $500$GeV this $t$-channel process is
dominant for Higgs masses in the range preferred by the indirect
electroweak precision data\cite{higgslimit2002} and remains an
important component for all Higgs masses at energies of the linear
collider. Tree-level computations of Higgs production are rather
well under control\cite{singlehzerwas,eewwhfusion}, including
interference of the $W$ fusion process with the Higgs-strahlung
process. Note however that the complexity of the process \nnht
precludes a full analytic result for the total cross section even
at tree-level, although the differential cross section can be cast
in a very compact form\cite{singlehzerwas,eewwhfusion}. Full
radiative corrections for Higgs-strahlung have been considered by
a number of groups\cite{eezhsmrc}, while a proper one-loop
treatment of the fusion process is still lacking despite the
importance of the process for the linear collider physics program.
Some recipes have been suggested\cite{kniehl-phys-rep,hwwapprox}
to include parts of the radiative corrections to the fusion
process but considering the domain of validity of these
approximations $\sqrt{s}, M_H \ll 2 m_t$, they are expected not to
be precise for the interesting range of Higgs masses(preferred by
the latest precision measurements\cite{higgslimit2002}) and next
collider energies. One-loop contribution to the $HWW$ vertex has
been considered on the basis that it might constitute a good
approximation for the fusion process\cite{nnhvienna}, but it rests
to see how well this approximation fares in comparison of the full
calculation. Very recently one-loop radiative corrections to this
process have been investigated within the minimal supersymmetric
model but again by only taking into account the contribution of
the fermions and sfermions to the $H/hWW$
vertex\cite{nnhvienna,Hahneetonnh}. It is the aim of this letter
to summarise the results of the full radiative corrections to
single Higgs production in \epemt, including both the fusion and
Higgs-strahlung processes in the \sm (Standard Model). We include
both the virtual and soft corrections as well as the hard photon
radiation. A longer paper will detail our computation and results
and will look into the issue of finding approximations to the full
result\footnote{Preliminary results have been presented at the
Workshop RADCOR2002\cite{eennhradcor2002}. At this meeting the
{\tt FIRCLA}\cite{Jegernnh} group exposed their plans and
techniques, different from ours concerning Feynman integration,
for tackling the calculation of this process. While finalising
this letter we also learnt of a calculation by A. Denner, S.
Dittmaier, M. Roth and M. Weber, in preparation.}.

A standard hand calculation using the usual  techniques could
hardly be attempted for such $2\ra 3$ processes at one-loop.
Considering the ever increasing power of computers, the
possibility of parallelisation and the fact that the whole
procedure of perturbation theory consists of algorithms that can
be directly translated on a computer it seems that most, if not
all, complex calculations in high-energy physics can be automated.
This is especially true for electroweak processes where various
scales and masses enter the calculations. \grcl\cite{grace-loop}
from which our results are derived is such a program.
\grc\cite{grace}, the tree-level component of the system, has been
tested and heavily used for tree-level cross sections up to
6-fermions in the final state\cite{graceee6f}. \grcl has been
exploited and checked thoroughly for a variety of $2\ra 2$
processes in the electroweak theory \cite{nlgfatpaper}. The system
which requires as input, a model file that describes all the
interaction vertices derived from a particular Lagrangian can
generate all the necessary Feynman graphs together with their
codes so that matrix elements can be generated before being
processed for the calculation of the cross section and  event
generation. For loop processes, there is a symbolic manipulation
stage (either {\tt FORM}\cite{form} or {\tt REDUCE}\cite{reduce} )
that handles all the Dirac and tensor algebra in $n$-dimension for
all the interference terms between tree-level and 1-loop diagrams
and automatically applies the Feynman trick for the propagator.
This is then passed to a module that contains two libraries for
the loop integration containing the {\tt FF} package\cite{ff} as
well as an in-house numerical code. The system together with the
one-loop renormalisation program is described in detail in
\cite{nlgfatpaper}. As far as the calculation of one-loop
processes is concerned, a series of powerful tests are implemented
in the code as described in \cite{nlgfatpaper} and as will be
presented below for \nnht.

\section{Tree-level results, setting-up the loop calculation}
Our input parameters for the calculation of \nnht are the
following. Throughout we expressed our results in terms of the
fine structure constant in the Thomson limit
$\alpha^{-1}=137.0359895$ and the $Z$ mass $M_Z=91.1876$GeV. Our
on-shell renormalisation program uses $M_W$ as input parameter,
nonetheless our numerical value of $M_W$ is derived through
$\Delta r$\cite{Hiokideltar}\footnote{We include NLO QCD
corrections and two-loop Higgs effects. We take
$\alpha_s(M_Z^2)=.118$ together with $G_\mu=1.16639\times
10^{-5}{\rm GeV}^{-2}$}. $M_W$ thus changes as a function of
$M_H$. For the  the lepton masses we take $m_e=0.510999$MeV,
$m_\mu=105.6584$MeV and $m_\tau=1.777$GeV. For the quark masses
beside the top mass $M_t=174$GeV, we take the set $M_u=M_d=58$MeV,
$M_s=92$MeV, $M_c=1.5$GeV and $M_b=4.7$GeV. With these values we
calculate $\Delta \alpha(M_Z)=0.059258$. With this we find for
example that $M_W=80.3767$GeV for $M_H=150$GeV and
$M_W=80.3158$GeV for $M_H=350$GeV. Especially for the
Higgs-strahlung subprocess we require a  $Z$-width. We have taken
a constant fixed $Z$-width, $\Gamma_Z=2.4952$GeV. Unless when
otherwise stated our results refer to the full \nnht, summing over
all three types of neutrinos with, for electron neutrinos, the
effect of interference between fusion and Higgs-strahlung.

We have checked that our tree-level results are in very good
agreement with those in \cite{singlehzerwas} after expressing them
in terms of $G_\mu$. However since we are considering the effect
of radiative corrections, within our scheme we prefer showing all
our results using $\alpha$. We will only comment on the $G_\mu$
scheme at the end of this letter.  We find for example for
$\sqrt{s}=500$GeV and $M_H=350$GeV that $\sigma_{{\rm tree, total
}}=4.606$fb at tree-level for the contribution of all three
neutrinos. In an attempt to separate the different contributions
to single Higgs production, we will refer to the $s$-channel as
given by $\sigma_s=3 \times \sigma (\epem \ra \nu_\mu
\bar{\nu}_\mu H)$. The bulk of this contribution is given by
$\sigma(\epem \ra ZH) \times B_{Z\ra {\rm inv.}}\;,\;B_{Z\ra {\rm
inv.}} \simeq 20\%$. We will define the $t$-channel as
$\sigma_t=\sigma_{{\rm total}}-\sigma_s$, this implicitly means
that the interference term is included in this contribution. These
definitions will be carried over to the one-loop case as well.

In Fig.~\ref{mh150s500} we have also included the tree-level cross
section. They are shown as a function of the centre-of-mass energy
for a light Higgs  of mass $M_H=150$GeV as well as a function of
the Higgs mass at a centre-of-mass of $500$GeV. All integration
over phase space are done with the help of {\tt BASES},
see\cite{grace}. These figures clearly show the importance of the
$t$-channel contribution pointed out in the introduction. For a
low Higgs mass of $150$GeV, although the $s$-channel still
dominates at $\sqrt{s}=300$GeV, very quickly at $\sqrt{s}=500$GeV
it is the $t$-channel that dominates. For the latter  energy as
the Higgs mass increases, the $t$-channel contribution drops much
quickly than the $s$-channel, where both merge around $M_H=380$GeV
to  $M_H=390$GeV, but very quickly around the $ZH$ threshold for
$M_H \sim 408$GeV, the $s$-channel drops precipitously leaving the
$t$-channel as the sole contribution for the whole process.

\begin{figure*}[htbp]
\begin{center}
\includegraphics[width=16cm,height=16cm]{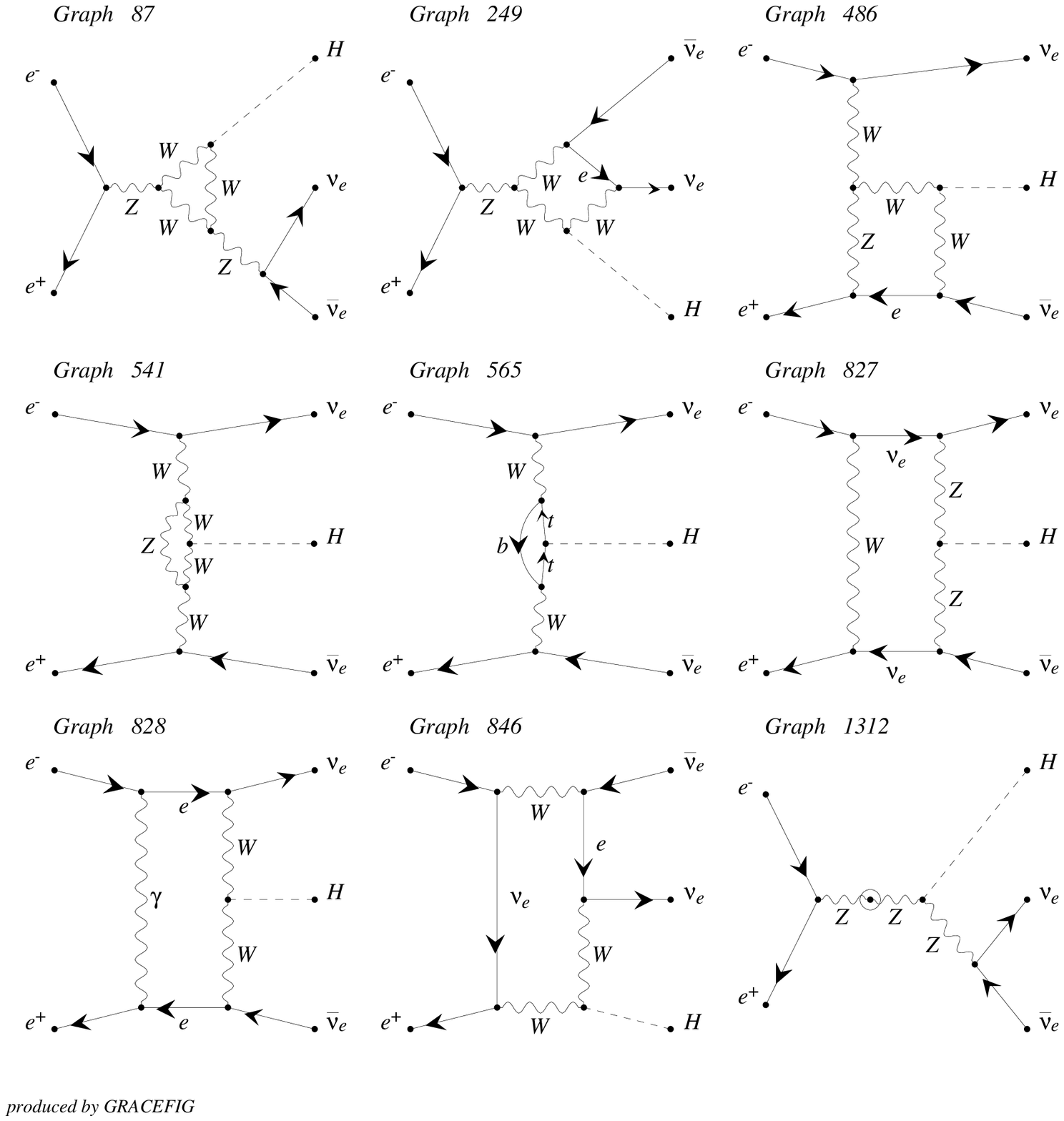}
\caption{\label{diagrams} {\em A small selection of different
classes of loop diagrams contributing to \nnht. We keep the same
graph numbering as that produced by the system. {\tt Graph 1312}
belongs to the corrections from self-energies, here both the
virtual and counterterm contributions are generated and counted as
one diagram. {\tt Graph 87} shows a vertex correction. Both graphs
can be considered as resonant Higgs-strahlung contributions.  {\tt
Graph 249} represents a box correction, it is a non resonant
contribution but applies also to the $\nu_\mu, \nu_\tau$ channels.
{\tt Graph 486} is also a box correction which is non resonant and
applies only to $\nu_e$.  {\tt Graph 541} and {\tt Graph 565} are
typical bosonic and fermionic corrections to the $WWH$ vertex for
the fusion process. {\tt Graph 846} shows a pentagon correction
that also applies to $\mu$ and $\tau$ neutrinos, this again can be
considered as a non-resonant contribution.  {\tt Graphs 827} and
{\tt 828} are pentagons that only contribute to \nnhet\/.}}
\end{center}
\end{figure*}

Neglecting all Goldstone-electron coupling (proportional to the
electron mass), one has at one-loop $249$ diagrams (for $\nu_e
\bar{\nu}_e H$, and $146$ for $\nu_\mu \bar{\nu}_\mu H$) including
$15$ pentagons (5-point functions) compared to only 2 diagrams at
tree-level, one for the fusion process and one for the
Higgs-strahlung. Keeping the electron Yukawa coupling one has a
total of $1350$ diagrams (98 pentagons corresponding to 5-point
functions) at one-loop. Although in running our program to derive
cross sections we only use the set of 249 diagrams, to perform our
extensive checks especially those of gauge-parameter independence,
at the level of the differential cross section, we keep the full
set of 1350 diagrams. It is impossible to show all the
contributing diagrams here. They may  be downloaded or visualised
at this location\cite{allfigseennh}. All these diagrams are
generated and drawn by {\tt gracefig} the Feynman diagrams
generator of \grc. A representative selection of diagrams is shown
in Fig.~\ref{diagrams}.

 The results of
the calculation are checked by performing three kinds of tests at
some random points in phase space. For these tests to be passed
one works in quadruple precision. We first check the ultraviolet
finiteness of the results. This test applies to the whole set  of
the virtual one-loop diagrams. In order to conduct this test we
regularise any infrared divergence by giving the photon a
fictitious mass (we set this at $\lambda=10^{-15}$GeV). In the
intermediate step of the symbolic calculation dealing with loop
integrals (in $n$-dimension), we extract the regulator constant
$C_{UV}=1/\varepsilon -\gamma_E+\log 4\pi$, $n=4-2 \varepsilon$
and treat this as a parameter. The ultraviolet finiteness test
gives a result that is stable over $30$ digits when one varies the
dimensional regularisation parameter $C_{UV}$. This parameter
could then be set to $0$ in further computation.  The test on the
infrared finiteness is performed by including both loop and
bremsstrahlung contributions and checking that there is no
dependence on the fictitious photon mass $\lambda$. We find
results that are stable over $23$ digits when varying $\lambda$.
An additional test concerns the bremsstrahlung part. It relates to
the independence in the parameter $k_c$ which is a soft photon cut
parameter that separates soft photon radiation  and the hard
photon performed  by the Monte-Carlo integration.

Gauge parameter independence of the result is performed through a
set of five gauge fixing parameters. For the latter a generalised
non-linear gauge fixing  condition\cite{nlg-generalised} has been
chosen.
\beqn
\label{fullnonlineargauge} {{\cal L}}_{GF}&=&-\frac{1}{\xi_W}
|(\partial_\mu\;-\;i e \tilde{\alpha} A_\mu\;-\;ig c_W
\tilde{\beta} Z_\mu) W^{\mu +} + \xi_W \frac{g}{2}(v
+\tilde{\delta} H +i \tilde{\kappa} \chi_3)\chi^{+}|^{2} \nonumber \\
& &\;-\frac{1}{2 \xi_Z} (\partial.Z + \xi_Z \frac{g}{ 2 c_W}
(v+\tilde\varepsilon H) \chi_3)^2 \;-\frac{1}{2 \xi_A} (\partial.A
)^2 \;.
\eeqn

The $\chi$ represent the Goldstone. We take the 't Hooft-Feynman
gauge with $\xi_W=\xi_Z=\xi_A=1$ so that no ``longitudinal" term
in the gauge propagators contributes. Not only this makes the
expressions much simpler and avoids unnecessary large
cancelations, but it also avoids the need for high tensor
structures in the loop integrals. The use of five parameters is
not redundant as often these parameters check complementary sets
of diagrams. For example the parameter $\tilde{\beta}$ is involved
in all diagrams containing the gauge $WWZ$ and their Goldstone
counterpart, whereas $\tilde{\alpha}$ checks $WW\gamma$ and
$\tilde{\delta}$ is implicitly present in $WWH$. For each
parameter of the set
$\zeta=(\tilde{\alpha},\tilde{\beta},\tilde{\delta},\tilde{\kappa},\tilde{\epsilon})$
the first check is made while freezing all other four parameters
to $0$. We have also made checks with two parameters non-zero. In
principle checking for $2$ or $3$ values of the gauge parameter
should be convincing enough. We in fact go one step further and
perform a complete gauge parameter independence. To achieve this
we generate for each non-linear gauge parameter $\zeta$, the
values of the loop correction to the total differential cross
section as well as the contribution of each one-loop diagram
contribution for the five values $\zeta=0,\pm 1, \pm 2$. The
one-loop diagram contribution from each loop graph $g$, is defined
as

\beqn
{\rm d}\sigma_g= {\rm d}\sigma_g(\zeta)=\Re \left({{\cal
T}}^{loop}_g\cdot {{\cal T}}^{tree\ \dagger}\right) \;.
\eeqn

\noi ${{\cal T}}^{tree}$ is the tree-level amplitude summed over
all tree-diagrams\footnote{Therefore the tree-level  amplitude
does not depend on any gauge parameter. For the process at hand
nonetheless, some individual tree diagrams depend on the gauge
parameter $\tilde{\delta}$ and $\tilde{\epsilon}$ giving extremely
small contributions proportional to the electron mass. Our
numerical procedure to isolate the gauge parameter dependence
detects these tiny variations, replacing ${{\cal T}}^{tree\
\dagger}$ by any tree-level diagram instead of the sum, one is
able to differentiate between a variation in ${\rm d}\sigma_g$ due
to the loop diagram or the residual one from ${{\cal T}}^{tree\
\dagger}$.}. ${{\cal T}}^{loop}_g$ is the one-loop amplitude
contribution of the one-loop diagram $g$. A rapid look at the
structure of the Feynman rules of the non-linear gauge leads one
to conclude that for \nnht each contribution is a polynomial of
(at most) third degree in the gauge parameter and thus, that each
contribution, ${\rm d}\sigma_g$ may be written as

\beqn
{\rm d}\sigma_g={\rm d}\sigma_g^{(0)}+\zeta {\rm d}\sigma_g^{(1)}
+\zeta^2 {\rm d}\sigma_g^{(2)} +\zeta^3 {\rm d}\sigma_g^{(3)} \;,
\eeqn

For each contribution ${\rm d}\sigma_g$, it is a straightforward
matter, given the values of ${\rm d}\sigma_g$ for the five input
$\zeta=0,\pm 1, \pm 2$, to reconstruct ${\rm
d}\sigma_g^{(0,1,2,3)}$. This is what we do. In fact for each set
of parameters we automatically pick up all those diagrams that
involve a dependence on the gauge parameter.
The number of diagrams in this set depends on the parameter
chosen. In some cases a very large number of diagrams is involved.
For the process at hand this occurs with the parameter $\deltat$
where about $500$ diagrams are involved in the check.

We then verify that the differential cross section is independent
of $\zeta$

\beqn
{\rm d}\sigma=\sum_g {\rm d}\sigma_g=\sum_g {\rm d}\sigma_g^{(0)}
\;,
\eeqn

and therefore that

\beqn
{\rm sum}_i=\frac{\sum_g{\rm d}\sigma_g^{(i)}}{{\rm Max}_g\bigl(
|{\rm d}\sigma_g^{(i)}|\bigr)}\;\; ,\;\; i=1,2,3 \;,
\eeqn
vanishes.

\begin{table}
\begin{center}
\begin{tabular}{c|c|c|c|c|}
 \cline{2-5}
& $\#$ graphs & ${\rm sum}_3$ & ${\rm sum}_2$ & ${\rm sum}_1$ \\
&&&& \\
\hline
  $\alphat$ & 149 & $-$ & $10^{-28}$ & $10^{-30}$ \\
  $\betat$  & 314 & $-$ & $10^{-31}$ &$10^{-23}$  \\
  $\deltat$ & 477 {{\em \small (1059)}} & $10^{-20}$ & $10^{-20}$& $10^{-26}$ \\
  $\kappat $ & 122 & $-$ & $10^{-23}$ & $10^{-23}$ \\
  $\epsilont$  & 128 {{\em \small (132)}} & $-$ & $10^{-21}$ & $10^{-30}$ \\ \hline
\end{tabular}
\caption{\label{tabchecknlg} {\em Numerical size of sum$_{i}$ for
each non-linear gauge parameter. '\# of graphs' means the number
of  graphs that contributes to each sum depending on the gauge
parameter, the number between bracket include the residual
dependence from ${{\cal T}}^{tree\ \dagger}$. $-$ means that no
diagram is involved.\/}}
\end{center}
\end{table}

As seen from Table~\ref{tabchecknlg} agreement within $20$ to $30$
digits is observed. This agreement gets better if one gives the
electron mass a higher value, say $1$GeV. The gauge parameter dependence check
not only tests the various components of the input file (correct
Feynman diagrams for example) but also the symbolic manipulation
part and most important of all the correctness of all the
reduction formulae and the proper implementation of all the
$N$-point functions. This is quite useful when one deals with
5-point functions as is the case at hand. Talking of parametric
integrals all tensor reductions are done following the standard
procedure and then passing the scalar integrals to the {\tt FF}
package\cite{ff} or to our own specially optimised routines when
photon exchange is involved. The pentagon integrals are expressed
in terms of boxes as is now standard\cite{fivetofour}, our
procedure is outlined in the Appendix. We work in the on-shell
renormalisation scheme closely following \cite{kyotorc}. Apart
from masses and couplings, renormalisation is also carried for the
fields. In particular  we also require the residues of the
renormalised propagators of all physical particles to be unity. As
known\cite{eezhsmrc}, this procedure leads to a (very sharp)
threshold singularity in the wave function of the Higgs at the
thresholds corresponding to $M_H=2 M_W,2 M_Z$. Solutions  to
smooth this behaviour\cite{threshold-sing}, like the inclusion of
the finite width of the $W$ and $Z$, do exist but we have not
implemented them yet in the present version of \grcl. Therefore
when scanning over $M_H$ it is sufficient to  avoid these regions
within $1$GeV around the thresholds.

As a separate check on our implementation we have also computed
$H\ra WW$ and $\epem \ra ZH$, after tuning our parameters we find
excellent agreement with the
literature\cite{eezhsmrc,kniehl-phys-rep}.

\section{Results}

\subsection{Full  ${{\cal O}}(\alpha)$ results}
The results we show here include all 3 neutrino species. We first
discuss the full ${{\cal O}}(\alpha)$ which includes the hard
bremsstrahlung part. The separation between the $s$-channel and
the  $t$-channel is done in the same way as with the separation
done at tree-level. It should be noted  that in the one-loop
diagrams  that contribute to $\epem \ra \nu_\mu \bar{\nu}_\mu H$,
and which we classify as $s$-channel, there are diagrams which can
not be deduced from the one-loop corrections to the $s$-channel $2
\ra 2$ process \eezh. {\tt Graph 249} in Fig.\ref{diagrams} is one
example. The relative correction for all three contributions is
defined as $ \delta_{{{\cal O}}(\alpha)} \equiv
\frac{\sigma_{O(\alpha)}}{\sigma_{tree}} - 1$ and will be referred
to as the full one-loop ${{\cal O}}(\alpha)$ correction.

\begin{figure*}[hbtp]
\begin{center}
\includegraphics[width=16cm,height=10cm]{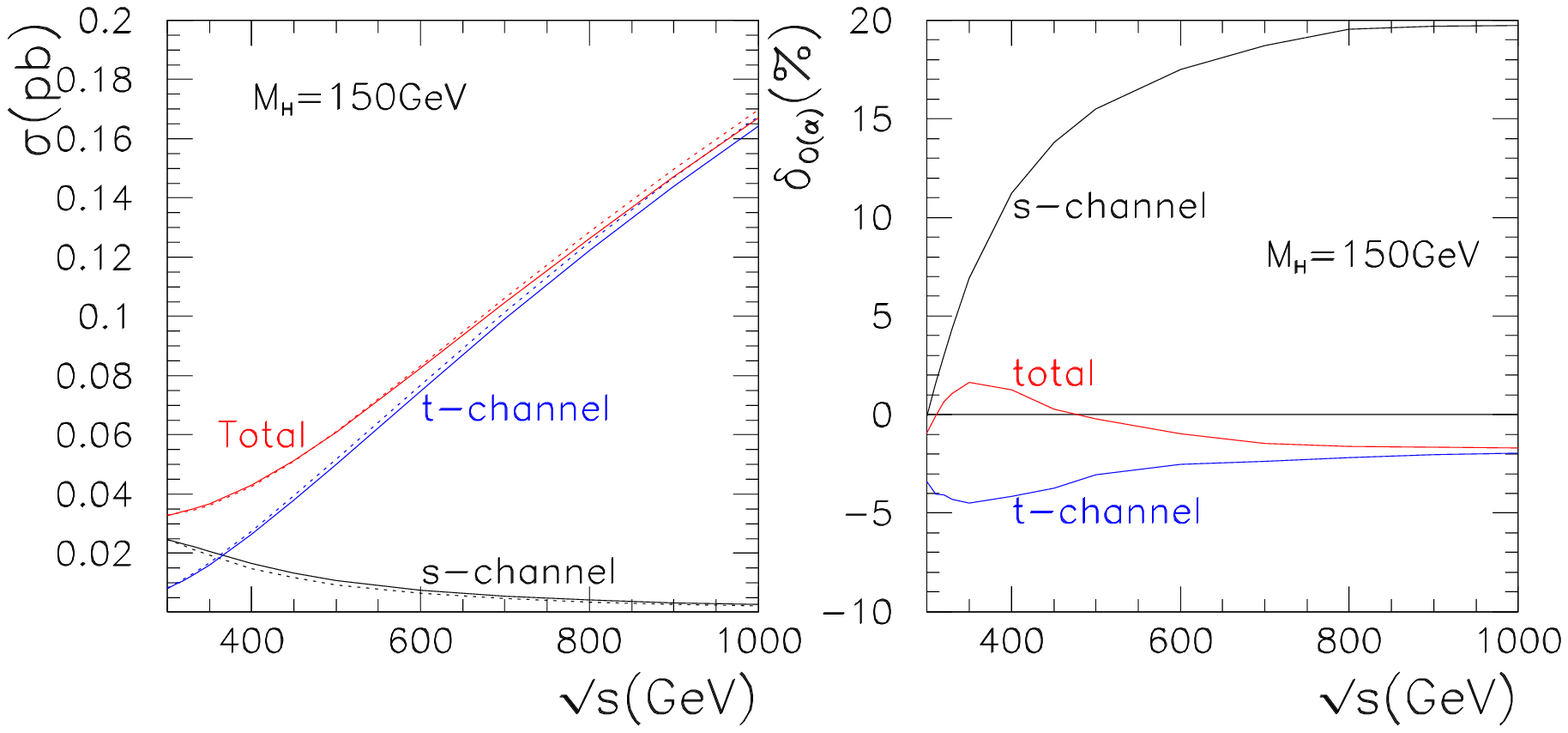}
\mbox{\hspace*{-1cm}
\mbox{\includegraphics[width=10cm,height=10cm]{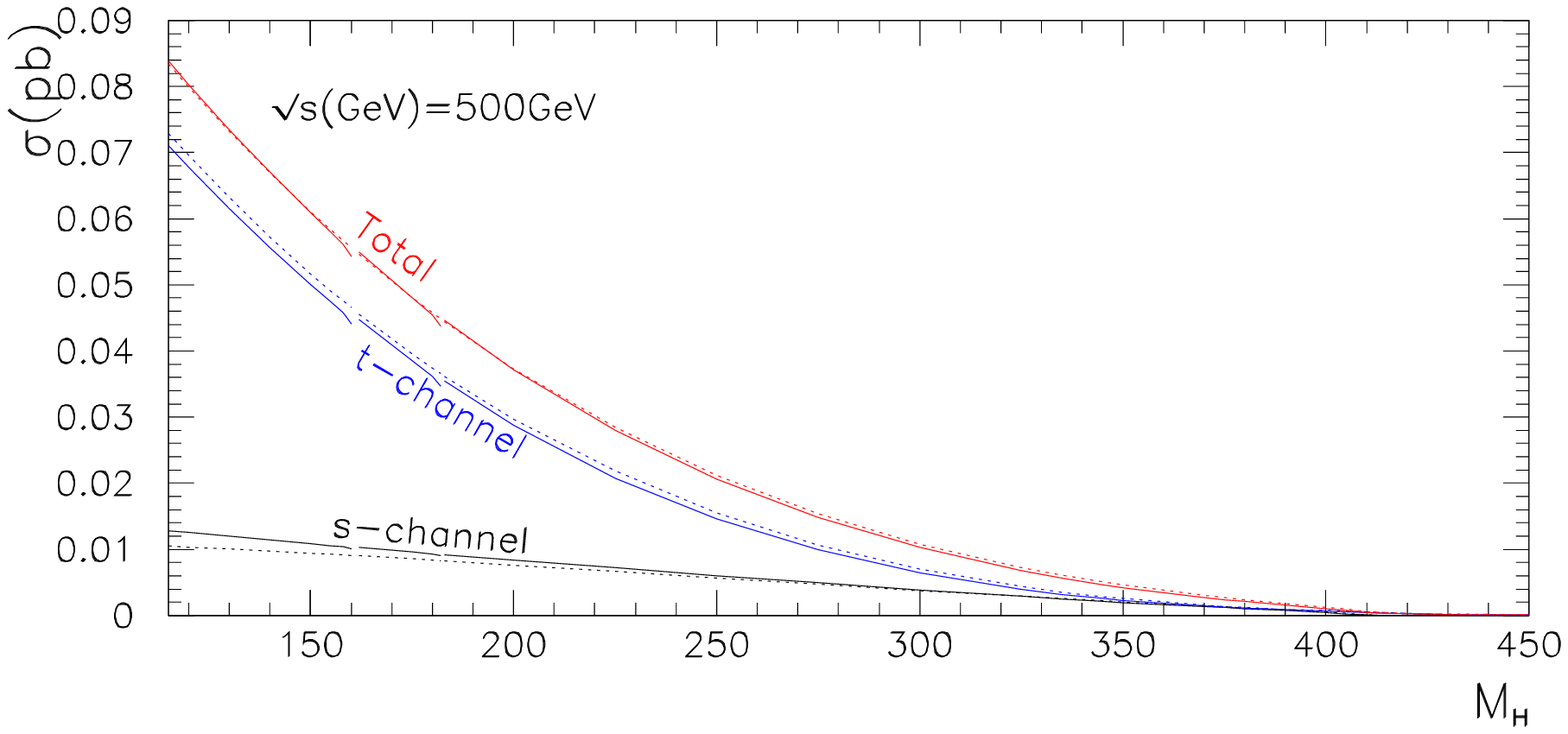}}
\raisebox{2.7cm}{ \mbox{ \hspace*{-6.5cm}
\includegraphics[width=6cm,height=7cm]{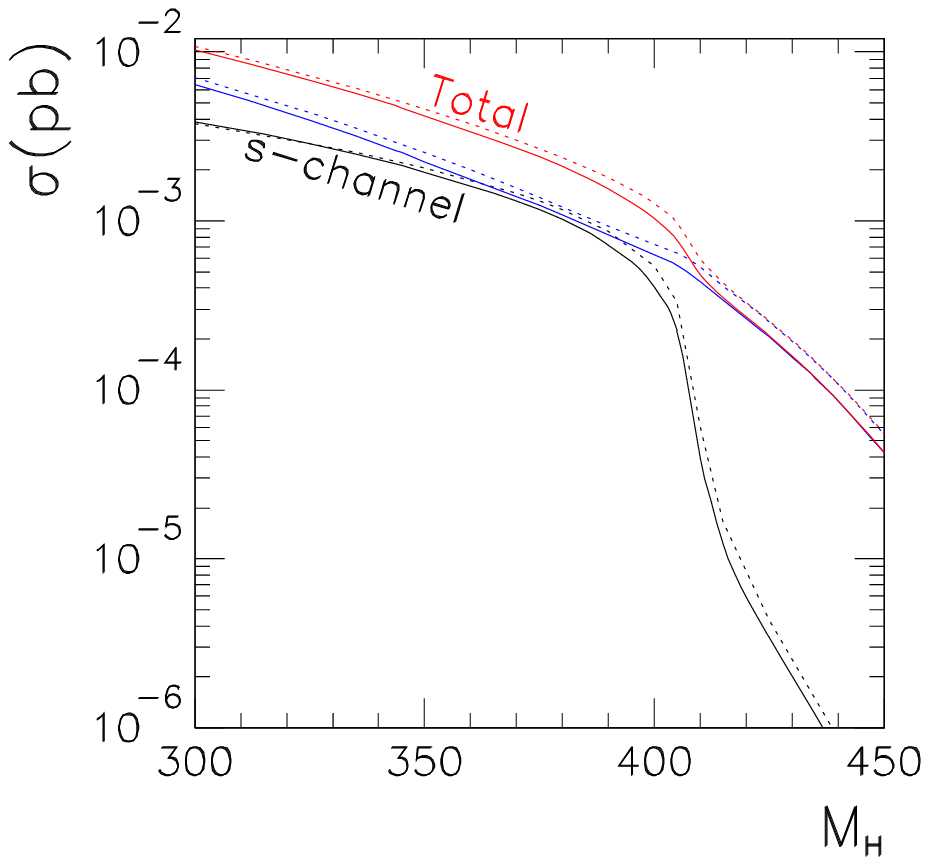}}}
\mbox{\hspace*{-.2cm}
\includegraphics[width=7cm,height=10cm]{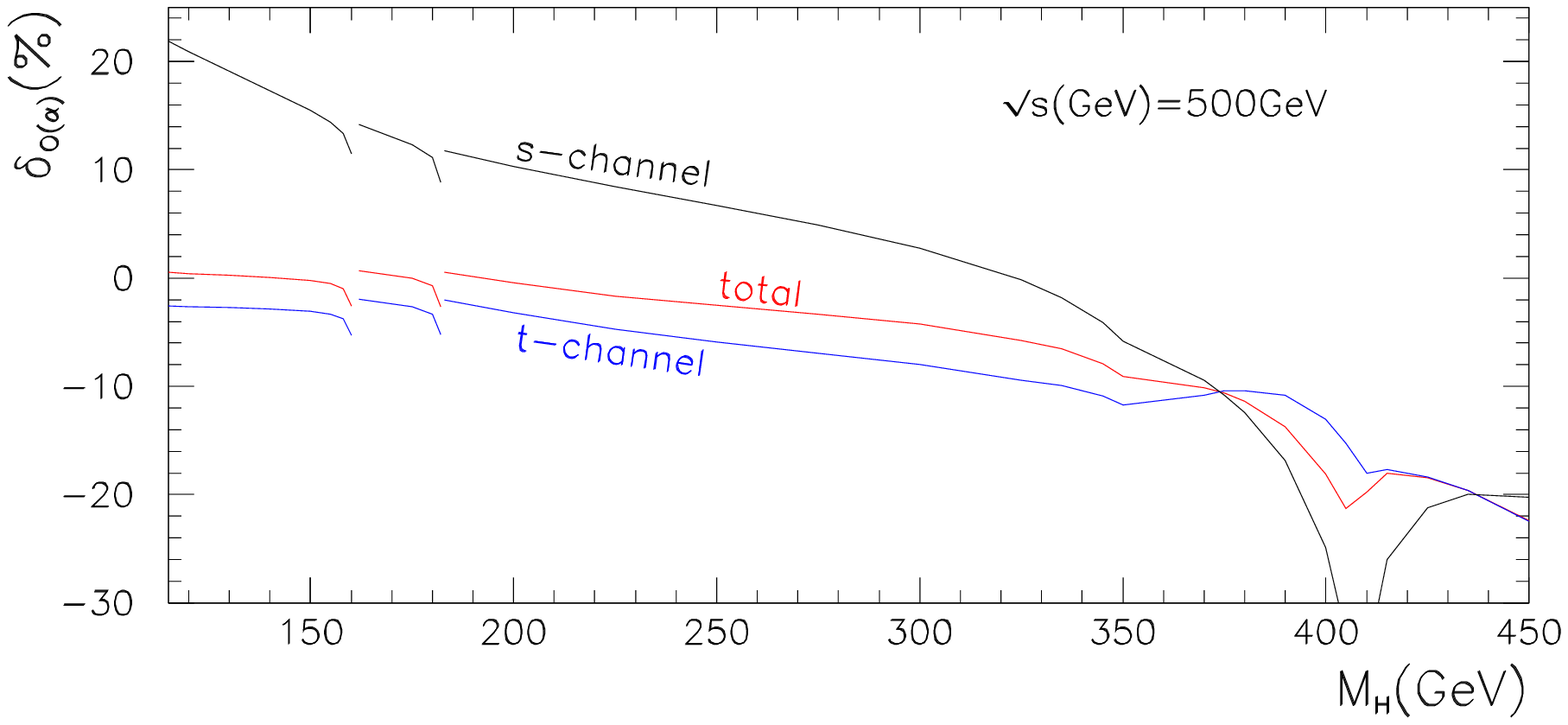}}}
\caption{\label{mh150s500} {\em The two figures in the first row
show  the cross section as a function of centre-of-mass energy for
a light Higgs  of mass $M_H=150$GeV. We show the $s$-channel,
$t$-channel and the sum of these (total) cross sections as defined
in the text. Both the tree-level (dashed lines) and the full
one-loop correction (full lines) are shown. In the second panel we
show the relative correction in per-cent. In the second row, the
dependence of the cross section as a function of the Higgs mass at
a centre-of-mass of $500$GeV is shown\/.}}
\end{center}
\end{figure*}

One important remark is that the overall correction in the
$s$-channel and $t$-channel are quite different. For a light Higgs
of mass $150$GeV the correction to the $s$-channel Higgs-strahlung
contribution is positive for practically all centre of mass
energies of the next linear collider, see Fig.~\ref{mh150s500}. It
rises rather sharply to reach about $+20\%$ for a centre-of mass
energy of 1TeV, however as will be argued below, the bulk of these
large corrections are due to virtual and real QED corrections.
Moreover in regions where these corrections are large, the
Higgs-strahlung contribution is rather small. On the other hand
the total correction in the $t$-channel for a small Higgs mass of
150GeV is negative throughout the range $\sqrt{s}=350$GeV to 1TeV,
and is almost constant past $\sqrt{s}=500$GeV, reaching about
$-2\%$. Combining these two contributions, we see that the full
correction to the whole process also remains small for a small
Higgs mass. In fact at $500$GeV it is almost at its lowest of
about $0.2\%$. This is an accidental cancellation between the
contributions of both the $t$-channel and the $s$-channel at this
energy. We have also studied the Higgs mass dependence of the
corrections. First of all note that all our results capture the
sharp spikes for $M_H=2 M_W, 2M_Z$, the top threshold is also
visible when we plot the relative corrections. For $500$GeV the
$s$-channel ${{\cal O}}(\alpha)$ correction decreases for higher
Higgs masses, eventually turning negative with a value $-6\%$ for
$M_H=350$GeV. It then drops rather sharply to reach as much as
$-30\%$ at the $ZH$ threshold, past which this cross section is
completely negligible. This behaviour is largely due to QED
corrections and is driven by the kinematics of the two-body $\epem
\ra ZH$.  At $500$GeV the correction in the $t$-channel
contribution remains negative for all Higgs masses that we
considered, {\it i.e.} in the range $115-450$GeV. It drops
steadily from about $-2\%$ for $M_H=115$GeV to about $-10\%$ at
$350$GeV close to the top pair threshold. It then increases up to
the $ZH$ threshold before dropping sharply around the $ZH$
threshold. Most of the large corrections are due to QED
corrections.

\subsection{Extraction the QED corrections}
As is known large QED corrections require a higher order
treatment. In order to quantify the effects of the genuine weak
corrections, one could try subtract these QED corrections. This
can be done rather easily for the $s$-channel contribution, where
the correction can be readily extracted from the electromagnetic
correction to the $eeZ$ vertex and the soft-photon bremsstrahlung
part. Indeed our computation produces at an intermediate stage the
result including the soft bremsstrahlung correction, that is
before the inclusion of hard photons. The cut on the photon
energy, $k_c$, has been taken sufficiently small, $k_c=0.1$GeV.
These corrections without hard bremsstrahlung include thus the QED
virtual and soft bremsstrahlung (which depend on $k_c$) as well as
the genuine weak correction to the process. For this $s$-channel
process the latter QED corrections are given by the universal soft
photon factor that leads to a relative correction

\beqn
\label{dqeduniv} \delta_{V+S}^{QED}=\frac{2
\alpha}{\pi}\left((L_e-1)\ln \frac{k_c}{E_b}+\frac{3}{4}L_e +
\frac{\pi^2}{6}-1 \right) \;,\; L_e=\ln(s/m_e^2) \;.
\eeqn
where $m_e$ is the electron mass and $E_b$ the beam energy
$s=4E_b^2$. Subtracting this contribution from our $k_c$ dependent
(numerical) result  reproduces the genuine weak correction,
$\delta_{W,s-channel}$.

To  quantify in an unambiguous way the effect of the weak
correction in the $t$-channel we have also subtracted this
universal factor $\delta_{V+S}^{QED}$ from the full ${{\cal
O}}(\alpha)$ correction. It can be shown that the leading
(infrared and collinear) contributions are given by the universal
factor, Eq.~\ref{dqeduniv}\cite{nlgfatpaper}. This procedure also
paves the way to a resummation of these large QED factors for the
full process which is conducive to a Monte-Carlo implementation as
could be done for instance through a QED parton
shower\cite{qedps}. This will be treated elsewhere. Coming back to
the weak corrections, we will denote by $\delta_{W,t-channel}$ and
$\delta_{W,total}$ the weak correction for the $t$-channel and the
full process based on the subtraction of the universal QED factor
in Eq.~\ref{dqeduniv}.

\begin{figure*}[hbtp]
\begin{center}
\mbox{\includegraphics[width=8cm,height=8cm]{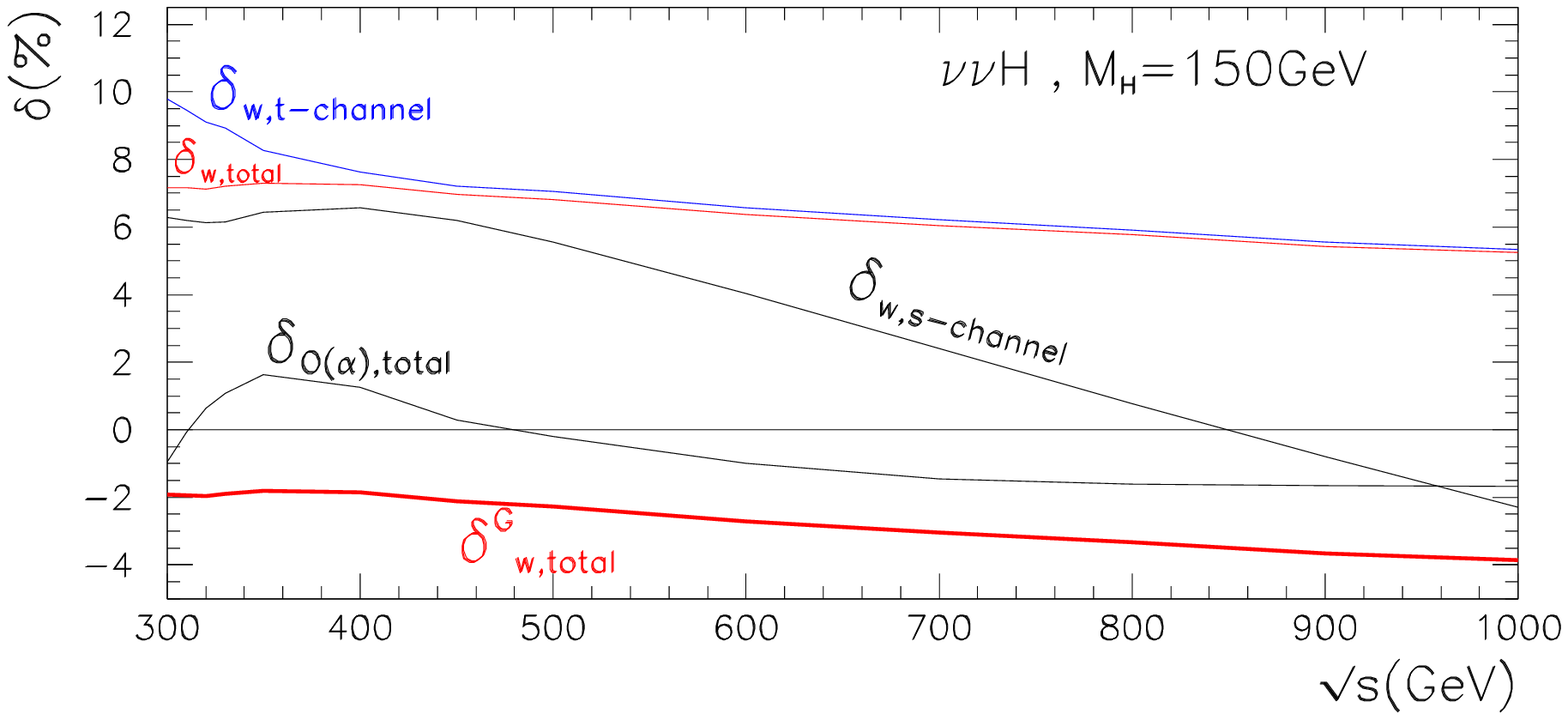}
\includegraphics[width=8cm,height=8cm]{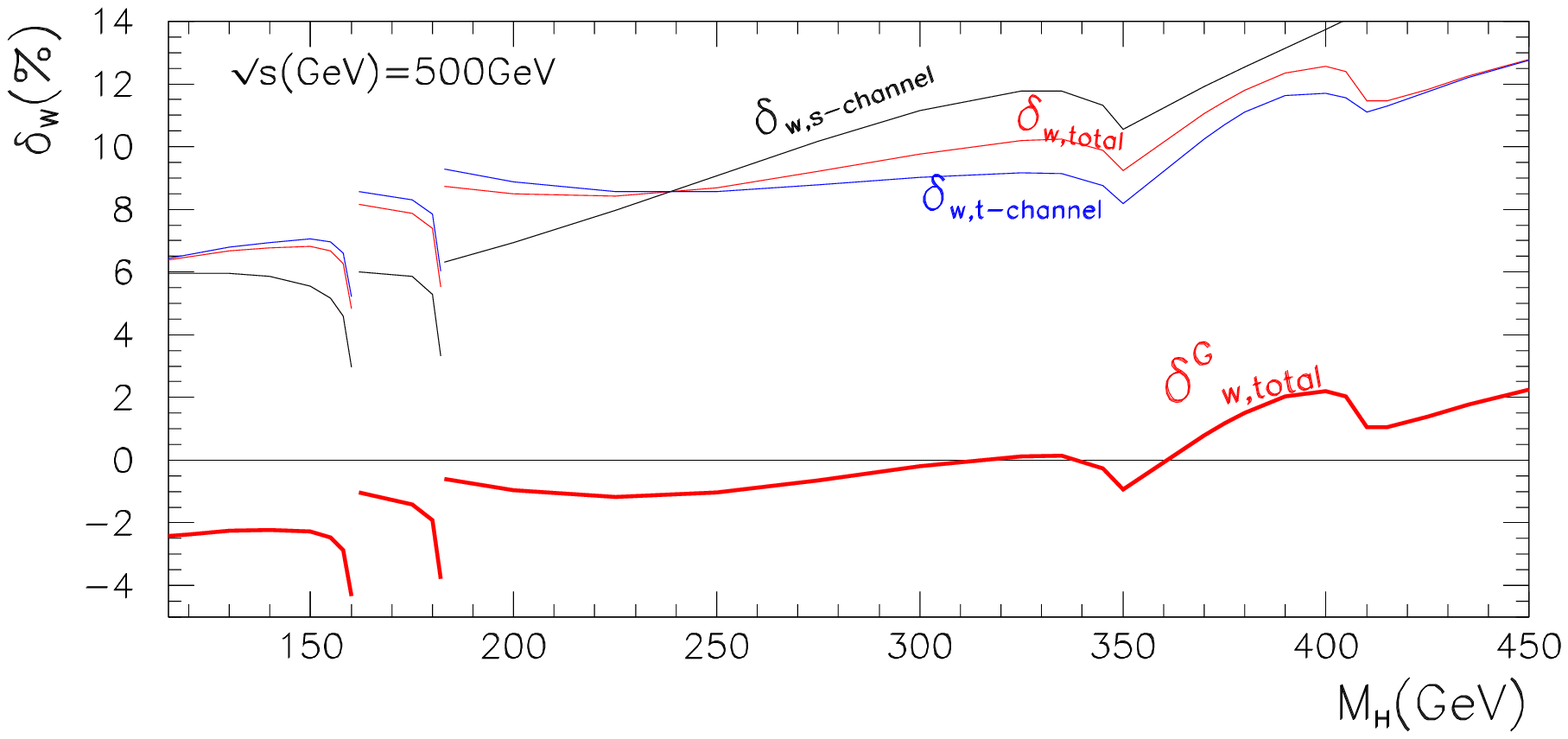}}
\caption{\label{figsdeltaw} {\em Relative weak corrections as
defined in the text, for the $t$-channel ($\delta_{W,t-channel}$),
$s$-channel ($\delta_{W,s-channel}$)and the whole process
($\delta_{W,total}$). We also show the full ${\cal O}(\alpha)$
correction for the whole process in the first panel. Also shown is
the weak correction for the full process expressed in the $G_\mu$
scheme ($\delta^G_{w,total}$), see text. \/}}
\end{center}
\end{figure*}

For a light Higgs mass ($M_H=150$GeV), the weak correction for the
$s$ and $t$-channels have a different behaviour as the energy
increases, see Fig.~\ref{figsdeltaw}. The former  varies  from
about $6\%$ at $\sqrt{s}=300$GeV to $-2.5\%$ at $\sqrt{s}=1$TeV.
Past $400$GeV where it dominates,   the weak correction to the
$t$-channel varies rather slowly from $7\%$ at $\sqrt{s}=400$GeV
to about $5\%$ at $\sqrt{s}=1$TeV. The dependence of the weak
corrections on the Higgs mass for a moderate centre-of-mass
energy, $\sqrt{s}=500$GeV,  reveals that up to the $ZH$ threshold
these corrections increase with the Higgs mass (most probably due
to  $M_H^2$ terms from  the Higgs self-coupling as in $\epem \ra
ZH$), apart from the clearly visible spikes at the $W,Z$ and the
top thresholds. Apart from the drop in the $t$-channel
contribution around the $ZH$ threshold, the weak correction in the
$t$-channel picks up again and as expected merges with the
correction to the full process.

\subsection{Expressing the weak corrections in terms of $G_\mu$}
Expressing the corrections in the $G_\mu$ scheme or in other words
had we expressed our tree-level results in terms of $G_\mu,$ thus
subtracting some universal weak corrections (essentially fermionic
contributions) affecting two-point functions, we can have a
quantitative measure of the non-universal weak radiative
correction specific to this process. We thus define for the
$s$-channel and $t$-channel contributions, these weak corrections
as $\delta_W^G=\delta_W-3 \Delta r$. Let us briefly summarise our
findings for $M_H=150$GeV where with our input $\Delta r$
contributes about $3\%$ (the leading Higgs mass dependence in
$\Delta r$ is logarithmic). For the full contribution with all
three neutrinos we find $\delta_W^G$ to be slowly varying (with
exactly the same ``slope" as $\delta_{W,total}$ in
Fig.~\ref{figsdeltaw}), from about $-2\%$ to about $-4\%$ in the
energy range from $\sqrt{s}=300$GeV to $\sqrt{s}=1$TeV. These
genuine weak corrections remain therefore well contained in the
full process, but in view of the precision of the \epemt machine
they must be taken into account. Applied to the $s$-channel with
$M_H=150$GeV, the corrections with $G_\mu$ as an input, are
moderate for energies up to $400$GeV but they quickly decrease
below about $-12\%$ at $1$TeV. Such behaviour had been observed in
\eezh\cite{ eezhsmrc}. This is another manifestation of the
failure of the $G_\mu$ scheme to properly describe the weak
corrections for such processes at high energies. For example it is
known that in \eezh$\;$ the contribution of boxes is important. We
do not attempt in this letter to make a thorough investigation of
the different loop contributions to \nnh, for example the
fermionic and bosonic contributions. We leave this to a further
study. This could be interesting in order to devise reliable
approximations based on a small subset compared to the large
number of contributions for such a complex process. For example,
very recently, the fermionic contributions and especially the
effect of the third generation have been investigated in
\cite{Hahneetonnh} and \cite{nnhvienna} with differing results. It
could be interesting to see how well these contributions can
reproduce the full result. We also do not report here on how the
distributions in the Higgs variables are affected by the radiative
corrections. We have briefly discussed this in a previous
note\cite{eennhradcor2002} and leave the full discussion for a
forthcoming paper.


\vspace*{1cm}

\section{Conclusions}
We have calculated with the help of \grcl the full radiative
corrections including hard photon radiation to the important Higgs
discovery channel at a future high energy \epemt machine, \nnht.
Apart from the usual checks on the ultraviolet and infra-red
finiteness of the result, we have performed tests on the gauge
parameter independence of the results. To this end we have relied
on a generalised non-linear gauge fixing condition where one has
control over five independent gauge parameters. For a light Higgs
of mass $150$GeV for energies ranging from $300$GeV to $1$TeV we
find a modest total ${\cal O}(\alpha)$ correction which is within
$\pm2\%$, being negligible at $500$GeV ($2$ per-mil). We have also
studied the Higgs mass dependence at $\sqrt{s}=500$GeV. For
example with $M_H=350$GeV we find a larger ${\cal O}(\alpha)$
negative correction of about $-10\%$. In order to quantify the
weak correction we have subtracted the universal QED virtual and
bremsstrahlung corrections. In the energy range  $\sqrt{s}=300$GeV
to $\sqrt{s}=1$TeV we find, for $M_H=150$GeV, that for the full
process the correction ranges from $+7\%$ to $+5\%$ when the
tree-level is expressed in terms of $\alpha$. Further
investigations and details on this important process are left to a
forthcoming publication.\\

\noi {\bf \large Acknowledgment} This work is part of a
collaboration between the {\tt GRACE} project in the Minami-Tateya
group and LAPTH. D. Perret-Gallix and Y. Kurihara deserve special
thanks for their contribution. This work was supported in part by
Japan Society for Promotion of Science under the Grant-in-Aid for
scientific Research B(N$^{{\rm o}}$. 14340081) and PICS 397 of the
French National Centre for Scientific Research.
\newcommand{\vect}[1]{\mathbf{#1}}

\renewcommand{\thesection}{\Alph{section}}
\setcounter{section}{0}
\renewcommand{\theequation}{\thesection . \arabic{equation}}
\setcounter{equation}{0}

\renewcommand{\thesection}{\Alph{section}}
\setcounter{section}{0}

\section{Appendix}

Five point functions are calculated as linear combinations of four
point functions\cite{fivetofour}. Our method is based on an
identity suitable for the Feynman parameter integration, which is
similar to the one described in\cite{nogueira5pt}.

A five point function is expressed as
\begin{eqnarray}
I_5 &=&   \int \frac{d^4 l}{(2 \pi)^4} \frac{N(l)}{D_0 D_1 D_2 D_3
D_4},
\end{eqnarray}
where \(l\) is the loop momentum and \(N(l)\) is a polynomial of
\(l^2\) and inner products of \(l\) with other four-vectors. The
denominators of propagators are defined as
\begin{eqnarray}
D_{0} &=& l^2 - m_{0}^{2} \;=\; l^2 + X_{0}, \nonumber\\
D_{i} &=& (l + r_{i})^{2} - m_{i}^{2} \;=\; l^2 + 2 l.r_i + X_{i},
\qquad i = 1, ..., 4.
                                            \label{di}
\end{eqnarray}

We take a set \(r_{i}\) (\(i = 1, ..., 4\)) of linearly
independent momenta. The latter form a basis for vectors in
4-dimensional space. Therefore with the Gram matrix $A_{ij} =
r_i.r_j$ one has the following identity
\begin{eqnarray}
\label{l2}
   g^{\mu\nu} = \sum_{i,j=1}^{4} r_{i}^\mu A_{ij}^{-1} r_{j}^\nu
   \Longrightarrow l^2 = \sum_{i,j=1}^{4} l.r_i A_{ij}^{-1} l.r_j.
\end{eqnarray}

Combining this identity with Eq.(\ref{di}) we obtain
\begin{eqnarray}
1 &=& \sum_{\alpha=0}^{4} [ a_{\alpha} + \sum_{i=1}^4 l.r_i
b_{\alpha,i} ]
                          D_{\alpha}, \label{sumab}
\label{oneab}
\end{eqnarray}
where
\begin{eqnarray}
a_{i} &=& \frac{1}{\Delta} \sum_{j=1}^4 A_{ij}^{-1} (X_j - X_0), \\
a_{0} &=&  \frac{4}{\Delta} - \sum_{i=1}^4 a_{i}, \\
b_{i,k} &=& - \frac{2}{\Delta} A_{ik}^{-1}, \\
b_{0,k} &=& - \sum_{i=1}^4 b_{i,k}, \\
\Delta &=& 4 X_{0} + \sum_{i,j} (X_i - X_0) A_{ij}^{-1} (X_j -
X_0) .
\end{eqnarray}
This immediately shows that the five point tensor integral can be
reduced to $5$ box integrals.

 Now we introduce the Feynman parameters. It is easy to see that
\begin{eqnarray}
I_5=\int \frac{d^4 l}{(2 \pi)^4} \int \prod_{\lambda=0}^4 d
x_{\lambda}
     \delta(1 - \sum_{\beta=0}^4 x_\beta)\;\;
\frac{N(l)}{D^4} \sum_{\alpha}
      \left( a_{\alpha} + \sum_{i=1}^4 l.r_i b_{\alpha,i} \right)
      \delta(x_\alpha).
\end{eqnarray}

Making a shift in the loop momentum , $l \rightarrow l-t\;\; t=
\sum_{i=1}^4 x_i r_i $,  so as to eliminate linear terms in the
loop momentum in $D$,  we obtain our reduction formula

\begin{eqnarray}
I_5&=&\int \frac{d^4 l}{(2 \pi)^4} \int \prod_{\lambda=0}^4 d
x_{\lambda}
     \delta(1 - \sum_{\beta=0}^4 x_\beta)\;\;
\frac{N(l-t)}{D^4} \sum_{\alpha=0}^4
      \left( \bar{a}_{\alpha} + \sum_{i=1}^4 l.r_i b_{\alpha,i} \right)
      \delta(x_\alpha) \nonumber \\
 \bar{a}_{0}&=&  a_0 - 2/\Delta \quad , \quad \bar{a}_{i} =
a_i\quad , \qquad  D=l^2 +\sum_{\alpha_=0}^{4} X_\alpha x_\alpha
-\sum_{i,j=1}^4 A_{ij} \;x_i x_j.
\end{eqnarray}

For the scalar pentagon, $N(l)=1$, only $\bar{a}_\alpha$ in the
previous equation contributes.


\begin{thebibliography}{99}

\bibitem{higgslimit2002}
M.W. Gr\"unewald, Plenary talk at the 31st ICHEP, Amsterdam,
  Netherlands,hep-ex/0210003. For updates see\\ {\tt
  http://lepewwg.web.cern.ch/LEPEWWG}.

\bibitem{singlehzerwas}
W. Kilian, M. Kr\"amer and P.M. Zerwas, Phys. Lett. {\bf B373}
(1996) 135.

\bibitem{eewwhfusion}
G.~Altarelli, B.~Mele and F.~Pitolli, Nucl.Phys. {\bf B287} (1987)
205. \\ R.N.
  Cahn Nucl.Phys.{\bf B255} (1985) 341 Erratum-ibid.{\bf B262} (1985)744. \\
  D.R.T. Jones and S.T. Petcov, Phys.Lett.{\bf B84} (1979) 440.\\ R.N. Cahn and
  S. Dawson,Phys.\ Lett.\ {\bf B136} (1984) 96; \\ G.L.\ Kane, W.W. Repko, and
  W.B.\ Rolnick, Phys. Lett. {\bf B148} (1984) 367; \\ B.A.\ Kniehl, Z.\ Phys.\
  {\bf C55} (1992) 605.\\ E.\ Boos, M.\ Sachwitz, H.\ Schreiber, and S.\
  Shichanin, Int.\ J.\ Mod.\ Phys.\ {\bf A10} (1995) 2067.

\bibitem{eezhsmrc}
A. Denner, J. K\"ublbeck, R. Mertig and M. B\"ohm, Z. Phys. {\bf
C56} (1992)
  261.\\ B.A. Kniehl, Z.~Phys. C55 (1992) 605.\\ See also, J. Fleischer and F.
  Jegerlehner, Nucl. Phys.~{\bf B216} (1983) 469.

\bibitem{kniehl-phys-rep}
B. A. Kniehl, Phys. Rep. {\bf 240} (1994) 211.

\bibitem{hwwapprox}
For a review see, B. A. Kniehl, Int.J.Mod.Phys. {\bf A17} (2002)
1457.

\bibitem{nnhvienna}
E. Eberl, W. Majerotto and V.C. Spanos, Phys. Lett. {\bf B538}
(2002) 35, {\it
  ibid} hep-ph/0210038.

\bibitem{Hahneetonnh}
T. Hahn, S. Heinemeyer and G. Weiglein, hep-ph/0211204.

\bibitem{eennhradcor2002}
G. B\'{e}langer, F. Boudjema, J. Fujimoto, T. Ishikawa, T. Kaneko,
K. Kato and
  Y.Shimizu, hep-ph/0211268, Proceedings of RADCOR 2002.

\bibitem{Jegernnh}
F.~Jegerlehner and O.~Tarasov, hep-ph/0212004, Proceedings of
RADCOR 2002..

\bibitem{grace-loop}
J. Fujimoto, T. Ishikawa, Y. Shimizu, K. Kato, N. Nakazawa and T.
Kaneko, Acta
  Phys. Polonica {\bf B28} (1997) 945.

\bibitem{grace}
T.Ishikawa, T.Kaneko, K.Kato, S.Kawabata, Y.Shimizu and K.Tanaka,
KEK Report
  92-19, 1993, The GRACE manual Ver. 1.0.

\bibitem{graceee6f}
F.~Yuasa, Y.~Kurihara and S.~Kawabata, \pl {\bf 414} (1997) 178.

\bibitem{nlgfatpaper}
G. B\'{e}langer, F. Boudjema, J. Fujimoto, T. Ishikawa, T. Kaneko,
K. Kato and
  Y.Shimizu, in preparation.

\bibitem{form}
J. A. M. Vermaseren:{\it New Features of FORM}; math-ph/0010025.

\bibitem{reduce}
{\em Reduce}, by A.C. Hearn: {\em Reduce User's Manual}, version
3.7, Rand.
  Corp. 1999.

\bibitem{ff}
G. J. van Oldenborgh , Comput. Phys. Commun. {\bf 58} (1991) 1.

\bibitem{Hiokideltar}
We use the code from Z. Hioki, see for example Z.Hioki, Zeit.
Phys. C49 (1991),
  287, see also Z. Hioki, Acta Phys.Polon. {\bf B27} (1996) 2573;
  hep-ph/9510269.

\bibitem{allfigseennh}
{\tt http://minami-home.kek.jp/eennh/grcfig-eennh.pdf} or \\
{\tt http://wwwlapp.in2p3.fr/$\sim$boudjema/eennh/alleennh.pdf},
the file is about 1.7Mb.

\bibitem{nlg-generalised}
F. Boudjema and E. Chopin, Z.Phys. {\bf C73} (1996) 85.

\bibitem{fivetofour}
D.B. Melrose, {\it Il Nuovo Cimento} {\bf 40A} (1965) 181. \\
W.L.~van Neerven
  and J.A.M.~Vermaseren, Phys.~Lett. {\bf 137} (1984) 241.

\bibitem{kyotorc}
K.~Aoki, Z.~Hioki, R.~Kawabe, M.~Konuma and T.~Muta, Suppl. Prog.
Theor. Phys.
  {\bf 73} (1982) 1.

\bibitem{threshold-sing}
T.~Bhattacharya and S.~Willenbrock, \pr {\bf D47} (1993) 469.\\
K.~Melnikov,
  M.~Spira and O.~Yakovlev, \zp {\bf C 64} (1994) 401. \\ B. A. Kniehl, C.P.
  Palisoc and A. Sirlin, \np {\bf B591} (2000) 296.

\bibitem{qedps}
Y~.Kurihara, J. Fujimoto, T. Munehisa and Y. Shimizu,
Prog.Theor.Phys. {\bf 96}
  (1996) 1223.\\ T. Munehisa, J. Fujimoto, Y. Kurihara and Y. Shimizu,
  Prog.Theor.Phys. {\bf 95} (1996) 375.

\bibitem{nogueira5pt}
P.Nogueira and J.C. Rom\~ao, Z.Phys. {\bf C60}, 757 (1993).

\end{thebibliography}
\end{document}